\documentclass{article}
\usepackage{graphicx}
\usepackage{amsbsy}
\begin{document}
\title{Early Universe: inflation and cosmological perturbations}
\author{David Langlois\\
{\small {\it APC (Astroparticules et Cosmologie),}}\\
{\small {\it UMR 7164 (CNRS, Universit\'e Paris 7, CEA, Observatoire de Paris)}}\\
{\small {\it  10, rue Alice Domon et L\'eonie Duquet, 75205 Paris Cedex 13, France}}}

\def\half{\frac{1}{2}}
\def\beq{\begin{equation}}
\def\eeq{\end{equation}}
\newcommand{\bea}{\begin{eqnarray}}
\newcommand{\eea}{\end{eqnarray}}
\def\Tdot#1{{{#1}^{\hbox{.}}}}
\def\Tddot#1{{{#1}^{\hbox{..}}}}
\def\p{\phi}
\newcommand{\dn}[2]{{\mathrm{d}^{{#1}}{{#2}}}}
\def\PX{P_{,X}}
\def\s{\sigma}

\def\tX{{\tilde X}}
\def\tG{{\tilde G}}
\def\tP{{\tilde P}}

\def\h{{\cal H}}

\def\d{\delta}
\def\B{{\bar B}}
\def\E{{\bar E}}
\def\V{{\cal V}}
\def\P{{\cal P}}
\def\R{{\cal R}}
\def\D{\nabla}

\def\L{{\rm L}}
\def\k{{\vec k}}
\def\x{{\vec x}}

\def\mP{m_P}

\def\bfphi{{\bf \phi}}

\def\bx{{\bf x}}
\def\bk{{\bf k}}
\def\PP{{\cal P}}

\maketitle

\begin{abstract}
After a brief summary of general relativity and cosmology, we present the basic concepts underlying inflation,
the currently best motivated models for  the early Universe. We describe the simplest inflation models, based on a single scalar field, and discuss how primordial cosmological perturbations are generated. 
We then review some recent developments concerning multi-field inflation models, in particular multi-field 
Dirac-Born-Infeld inflation. 
\end{abstract}

\section{Introduction}

Inflation, i.e. a phase of accelerated expansion, has now become a standard paradigm to describe the physics of the very early universe. Although many models of inflation have now been ruled out by observations, many remain
compatible with the present data and the nature of the field(s) responsible for inflation is still an open question. As more and more precise cosmological data will continue to accumulate in the coming years, one can envisage the 
fascinating possibility to learn from the observations some crucial clues about the fundamental physics at work in the very early Universe. 

In the first part of this contribution, we present a few basic results from general relativity and from standard 
cosmology.
The second part is devoted to the simplest models of inflation and to the computation of 
the  cosmological perturbations that these models generate, which is crucial for confrontation with cosmological observations. These first two parts are mainly based on the pedagogical introduction \cite{cargese} where the reader will find more details and references.

In the third  part, we go beyond the simplest models by extending our analysis to models of inflation involving several scalar fields. We show how the standard results are modified in this context and discuss various models which have attracted a lot of attention during the last years. 
We present  recent results for very general multi-field inflationary models, allowing for non-standard kinetic terms. This generalization is  motivated by efforts  to connect  string theory and inflation and we focus our attention on multi-field DBI (Dirac-Born-Infeld) inflation.  

\section{A few elements on general relativity and cosmology}
\subsection{General relativity}
The standard model of modern cosmology is based on Einstein's theory of general relativity. 
Without entering into details, which can 
be found in  standard textbooks on general relativity, let us recall a few useful notions.
In the framework of general relativity, the spacetime geometry is defined by 
a {\it metric}, a symmetric tensor with two indices, whose 
components in a coordinate system $\{x^\mu\}$ ($\mu=0,1,2,3$)  
will be denoted  $g_{\mu\nu}$.
The square of the ``distance'' between two neighbouring points of spacetime 
 is 
given by the expression 
\beq
ds^2=g_{\mu\nu}dx^\mu dx^\nu.
\eeq
We will use the signature  $(-,+,+,+)$.

It is convenient to define a {\it covariant derivative} associated to this metric, 
denoted $\D_\mu$, whose action on a tensor with, for example,
 one covariant index
and one contravariant index will be given by 
\beq
\D_\lambda T^{\mu}_{\hskip 2mm \nu}=\partial_\lambda T^{\mu}_{\hskip 2mm \nu}
+\Gamma^\mu_{\lambda\sigma}T^{\sigma}_{\hskip 2mm \nu}
- \Gamma^\sigma_{\lambda\nu}T^{\mu}_{\hskip 2mm \sigma}
\eeq
(a similar term must be added for each additional covariant or 
contravariant index),
where the  $\Gamma$ are the  {\it Christoffel symbols} (they are not 
tensors), defined by 
\beq
\Gamma^\lambda_{\mu\nu}={1\over 2}g^{\lambda\sigma}\left(\partial_\mu
g_{\sigma\nu}+\partial_{\nu}g_{\mu\sigma}-\partial_\sigma g_{\mu\nu}
\right).
\label{christoffel}
\eeq
We have used the notation $g^{\mu\nu}$ which corresponds, for the 
metric (and only for the metric), to the 
inverse of $g_{\mu\nu}$ in a  matricial sense, i.e. 
$g_{\mu\sigma}g^{\sigma\nu}=\delta_\mu^\nu$.

The ``curvature'' of spacetime is characterized by the {\it Riemann}
tensor, whose components can be expressed in terms of the 
Christoffel symbols according to the expression
\beq
R_{\lambda\mu\nu}^{\hskip 6mm \rho}=
\partial_\mu \Gamma_{\lambda\nu}^\rho
-\partial_\lambda \Gamma_{\mu\nu}^\rho+
\Gamma_{\lambda\nu}^\sigma\Gamma_{\sigma\mu}^\rho
- \Gamma_{\mu\nu}^\sigma\Gamma_{\sigma\lambda}^\rho.
\eeq

{\it Einstein's equations} relate the spacetime geometry to its matter content.
The geometry appears in Einstein's equations via the {\it Ricci tensor},
defined by 
\beq
R_{\mu\nu}=R_{\mu\sigma\nu}^{\hskip 6mm \sigma},
\eeq
and the  {\it scalar curvature}, which is the trace of the Ricci
tensor, i.e.
\beq
R=g^{\mu\nu}R_{\mu\nu}.
\eeq
The matter enters Einstein's equations via the {\it energy-momentum 
tensor}, denoted $T_{\mu\nu}$, whose time/time component 
corresponds to the energy density, the time/space components 
to the momentum density and the space/space component to the 
stress tensor. 
Einstein's equations then read
\beq
G_{\mu\nu}\equiv R_{\mu\nu}-{1\over 2}R\,  g_{\mu\nu}=8\pi G\,  T_{\mu\nu},
\label{einstein}
\eeq
where the tensor  $G_{\mu\nu}$ is called the  {\it Einstein tensor}.
Since, by construction, the  Einstein tensor satisfies the identity 
$\D_\mu G^{\mu}_{\, \nu}=0$,
any energy-momentum on the right-hand side of  Einstein's equation 
 must necessarily satisfy  the relation 
\beq
\D_\mu T^{\mu}_{\, \nu}=0, 
\label{DT0}
\eeq
which can be interpreted as a generalization, in the context of 
a curved spacetime, of the familiar conservation laws for energy and 
momentum.

Einstein's equations can also be obtained from a variational principle. The 
corresponding action reads
\beq
{\cal S}={1\over 16\pi G}\int d^4x\sqrt{-g}\left(R-2\Lambda\right)
+\int d^4x \sqrt{-g} {\cal L}_{mat}.
\eeq
One can check that the variation of this action with respect to the metric 
 $g_{\mu\nu}$ , upon using the  definition 
\beq
T^{\mu\nu}={2\over \sqrt{-g}}{\delta\left(\sqrt{-g}{\cal L}_{mat}\right)
\over{\delta g_{\mu\nu}}},
\label{def_Tmunu}
\eeq
indeed gives Einstein's equations 
\beq
G_{\mu\nu}+\Lambda g_{\mu\nu}=8\pi G\,  T_{\mu\nu}.
\eeq
This is a slight generalization of Einstein's equations (\ref{einstein}) that 
includes a {\it cosmological constant} $\Lambda$. 
It is worth noticing that the 
cosmological constant can also be interpreted as a particular 
energy-momentum tensor of the form
 $T_{\mu\nu}=-(8\pi G)^{-1}\Lambda g_{\mu\nu}$.

\subsection{Standard cosmology}

Let us now present briefly the tenets of modern cosmology. 
They are based on Einstein's equations  and on a few 
hypotheses concerning spacetime and its matter content. The most important one, 
so far confirmed by observations on large scales, is that our universe is approximately homogeneous and 
isotropic. 
Geometries that are strictly homogeneous and isotropic are described by  the so-called 
{\it  Robertson-Walker} metrics, which read in an appropriate coordinate 
system
\beq
ds^2=-dt^2+a^2(t)\left[{dr^2\over{1-\kappa r^2}}+
r^2\left(d\theta^2+\sin^2\theta d\phi^2\right)\right],
\label{RW}
\eeq
with $\kappa=0,-1,1$ depending on the curvature of spatial 
hypersurfaces: respectively flat, elliptic or hyperbolic.

The matter content compatible with the spacetime symmetries of homogeneity
and isotropy is necessarily described by an energy-momentum tensor of the 
form (in the same coordinate system as for the metric  (\ref{RW})),
\beq
T^{\mu}_{\, \nu}={\rm Diag}\left(-\rho(t), p(t),  p(t), p(t)\right), 
\eeq
where  $\rho$ corresponds to an energy density and $p$ to a pressure.

Substituting the  Robertson-Walker metric (\ref{RW}) 
in Einstein's equations (\ref{einstein}), one gets the 
 {\it  Friedmann-Lema\^\i tre equations}:
\begin{eqnarray}
\left({\dot a\over a}\right)^2 &=& {8\pi G\rho\over 3}- {\kappa\over a^2},
\label{friedmann}
\\
{\ddot a\over a} &=& -{4\pi G\over 3}\left(\rho+3 p\right).
\label{friedmann2}
\end{eqnarray}
An immediate consequence of these two equations is the {\it continuity 
equation}
\beq
\dot \rho+3H\left(\rho+p\right)=0,
\label{conservation}
\eeq
where $H\equiv \dot a/a$ is the {\it Hubble parameter}.
The continuity equation 
 can be also obtained directly from the energy-momentum conservation 
 $\D_\mu
T^{\mu}_{\, \nu}=0$.

In order to determine the cosmological time evolution, it is easier to combine
(\ref{friedmann}) with  (\ref{conservation}). 
Let us assume an equation of state for the cosmological matter of the form
$p=w\rho$ with
$w$ constant, which includes the two main types of matter that play an 
important r\^ole in cosmology, namely a gas of relativistic particles, with $w=1/3$, 
and non-relativistic matter, with $w\simeq 0$.
In these cases, the conservation equation  (\ref{conservation}) 
can be integrated to give
\beq
\rho\propto a^{-3(1+w)}.
\eeq
Substituting in (\ref{friedmann}), one finds,  for  $\kappa=0$,  
\beq
3{\dot a^2\over a^2}=8\pi G\rho_0\left(a\over a_0\right)^{-3(1+w)}, 
\eeq
where, by convention, the subscript '0' stands for {\it present}
quantities.
This implies for the 
evolution of the scale factor
\beq
a(t)\propto t^{\frac{2}{3(1+w)}}
\eeq
which thus gives $a(t)\propto t^{2/3}$ in a universe dominated by  non-relativistic matter
and $a(t)\propto t^{1/2}$ in a universe dominated by  radiation.

One can also mention
 the case of a {\it cosmological constant}, which corresponds
to an equation of state $w=-1$ and thus implies 
an exponential evolution for the scale
factor
\beq
a(t)\propto \exp(Ht).
\eeq

More generally, when  several types of matter coexist with respectively 
 $p_{(i)}=w_{(i)}\rho_{(i)}$, it is convenient to introduce the 
dimensionless parameters 
\beq
\Omega_{(i)}={8\pi G \rho_0^{(i)}\over 3 H_0^2},
\eeq
which express the {\it present} ratio of  the energy density of the various
 species with   the so-called {\it critical energy density} 
$\rho_{crit}= 3 H_0^2/(8\pi G)$, which corresponds to the total energy
density for a flat universe.

One can then rewrite the first 
 Friedmann equation  (\ref{friedmann}) as 
\beq
\left({H\over H_0}\right)^2=\sum_i\Omega_{(i)}
\left(a\over a_0\right)^{-3(1+w_{(i)})}+\Omega_\kappa
\left(a\over a_0\right)^{-2},
\eeq
with $\Omega_\kappa=-\kappa/a_0^2H_0^2$, which implies that the 
cosmological parameters must satisfy the consistency relation
\beq
\sum_i \Omega_{(i)}+\Omega_\kappa=1.
\eeq
As for the second Friedmann equation (\ref{friedmann2}), it implies
\beq
{\ddot a_0\over a_0 H_0^2}=-{1\over 2}\sum_i \Omega_{(i)}(1+w_{(i)}).
\eeq
Cosmological observations yield for the various parameters (see e.g. \cite{wmap5})
\begin{itemize}
\item Baryons: $\Omega_b\simeq 0.05$,

\item Dark matter: $\Omega_d\simeq 0.23$,

\item Dark energy (compatible with a cosmological constant): 
$\Omega_{\Lambda}\simeq 0.72$,

\item Photons: $\Omega_\gamma\simeq 5\times 10^{-5} $.
\end{itemize}
Observations have not detected so far any deviation from flatness. 
Radiation is very subdominant today but extrapolating 
backwards in time, radiation was dominant in the past 
  since its energy density scales 
as $\rho_\gamma\propto a^{-4}$ in contrast with non-relativistic matter 
($\rho_m\propto a^{-3}$). 
Moreover, the matter content today seems to be dominated by some dark energy with 
a present equation of state very close to that 
of a cosmological constant ($w_\Lambda=-1$), which means that 
our universe is currently accelerating.

To go beyond a purely geometrical description of cosmology,
it is useful to apply thermodynamics to  the matter content of the 
universe. One can then  define a temperature $T$ for the cosmological 
photons, not only when they are strongly interacting with ordinary matter
but also after they have decoupled because, 
with the  expansion, the thermal distribution for the gas of 
photons is unchanged  except for a global rescaling of the 
temperature so that $T$ essentially evolves as 
\beq
T(t)\propto {1\over a(t)}.
\eeq
This means that, going  backwards in time, the universe was much hotter 
than today. This is the essence  of the  Big Bang scenario.

 As the universe evolves,  the reaction rates between the various 
species  are modified. A detailed analysis of these changes allows  
to reconstruct the past thermal history 
of the universe. 
Two events in  particular play an essential  r\^ole because of 
their observational consequences:
\begin{itemize}

\item Primordial nucleosynthesis

Nucleosynthesis occured at a temperature 
around $0.1$ MeV, when the average kinetic energy  
became  sufficiently low so that nuclear binding was possible. Protons 
and neutrons could then combine, which lead to the production of 
light elements, such that Helium, Deuterium, Lithium, etc... Within 
the observational uncertainties, this 
scenario is  remarkably confirmed  by the present measurements. 

\item Decoupling of baryons and photons (or last scattering)

A more recent event is the so-called ``recombination'' of nuclei and 
electrons to form atoms. This occured  at a temperature of the order of 
the eV. Free electrons thus almost disappeared, which entailed an effective
decoupling of the cosmological photons from ordinary matter. What we 
see today as the Cosmic Microwave Background (CMB) is made of the fossil 
photons, which interacted  for the last time with matter at the time of recombination. 
The CMB represents a remarkable  observational 
tool for analysing the perturbations of the  early universe, as well as
for measuring the cosmological parameters introduced above.

\end{itemize}

\subsection{Puzzles of the standard Big Bang scenario}
The standard Big Bang scenario has encountered remarkable successes, in 
particular with the nucleosynthesis scenario and the prediction of the CMB, 
and it remains today a cornerstone in our understanding of the present 
and past universe.
However, a few intriguing facts  remain unexplained in the basic  Big Bang model 
and seem to necessitate a larger 
framework. Concerning the present matter state of the Universe, 
we need of course to understand the nature of dark matter and of dark energy.
In addition to these two crucial questions, several properties of our Universe 
are problematic in the perspective of its evolution. 
We now review these problems.

\begin{itemize}

\item Homogeneity problem

A first question is why the approximation of homogeneity and isotropy 
turns out to be so good. Indeed, inhomogeneities are 
unstable, because of gravitation, and they tend to grow with time. 
It can be verified for instance with the CMB that inhomogeneities were 
much smaller at the last scattering epoch than today. One thus expects that 
these homogeneities were still smaller further back in time.
How to explain a universe so smooth in its past ?

\item Flatness problem

Another puzzle is the (spatial) flatness of our universe. Indeed, 
Friedmann's equation (\ref{friedmann}) implies
\beq
\Omega-1\equiv {8\pi G \rho\over 3 H^2}-1={\kappa\over a^2 H^2}.
\eeq
In standard cosmology, the scale factor behaves like
$a\sim t^q$ with $q<1$  ($q=1/2$ for radiation and $q=2/3$ for 
non-relativistic  matter).
 As a consequence,  $(aH)^{-2}$  grows with time and 
  $|\Omega-1|$ must thus  diverge with time. Therefore, in the 
context of the standard model, the quasi-flatness observed today 
requires an extreme fine-tuning of $\Omega$ near $1$ in the early 
 universe.

\item Horizon problem


One of the most fundamental problems in standard cosmology is certainly
the {\it horizon problem}.
The (particle) {\it horizon} is the maximal distance that can be covered 
by a light ray. 
For a light-like radial trajectory 
 $dr=a(t) dt$ and   the horizon is thus given by 
\beq
d_{H}(t)= a(t)\int_{t_i}^t{dt'\over a(t')}=a(t){t^{1-q}-t_i^{1-q}\over 1-q}, 
\eeq
where the last equality is obtained by assuming $a(t)\sim t^q$ and $t_i$ is
some initial time.

In standard cosmology ($q<1$), the integral converges in the limit 
 $t_i=0$ and the horizon has a finite size, of the order of the 
so-called Hubble radius $H^{-1}$:
\beq
d_H(t)={q\over 1-q}H^{-1}.
\eeq 
 
It also useful to consider 
 the {\it comoving Hubble radius}, $(aH)^{-1}$, which represents
the fraction of comoving space in causal contact.
One finds that it  {\it grows} with time, which means that the 
{\it fraction of the universe in causal contact increases with time} in 
the context of standard cosmology.
But the CMB tells us that the Universe was quasi-homogeneous 
at the time of last scattering on a scale encompassing 
 many regions a priori causally independent.
How to explain this ? 
\end{itemize}

A solution to the horizon problem and to the other puzzles 
is provided by the inflationary scenario, which 
we will examine in the next section. The basic idea
 is to invert the behaviour of the 
comoving Hubble radius, that is to make it {\it decrease} sufficiently in the very 
early universe. 
The corresponding condition is that 
\beq
\ddot a>0,
\eeq
i.e. that the Universe must undergo a {\it phase of acceleration}.

\section{Single field inflation}
The simplest inflationary models are based on  a single scalar field $\phi$ governed by an 
action of the form
\beq
S=\int d^4x\sqrt{-g}\left(-{1\over 2}\partial^\mu\phi\partial_\mu\phi
-V(\phi)\right),
\eeq
where $V(\phi)$ is the potential for the scalar field.
The corresponding energy-momentum tensor is given by  
\beq
T_{\mu\nu}=\partial_\mu\phi\partial_\nu\phi-g_{\mu\nu}
\left({1\over 2}\partial^\sigma\phi\, \partial_\sigma\phi
+V(\phi)\right).
\label{Tscalarfield}
\eeq

\subsection{Homogeneous evolution}
In a spatially flat FLRW (Friedmann-Lemaitre-Robertson-Walker) spacetime, with metric
\beq
ds^2=-dt^2+a^2(t) d{\vec x}^2,
\eeq
 the energy-momentum tensor reduces to the perfect fluid form
with  energy density and pressure given respectively by
\beq
\rho=-T_0^0={1\over 2}\dot\phi^2+V(\phi),\quad 
p={1\over 2}\dot\phi^2-V(\phi).
\eeq
The equation of motion for the scalar field is 
\beq
\label{KG}
\ddot\phi+3H\dot \phi+V'=0. 
\eeq
and the evolution of the scale factor is governed by  Friedmann's equations
\begin{eqnarray}
H^2={8\pi G\over 3}\left({1\over 2}\dot\phi^2+V(\phi)\right), \qquad  \dot H=-4\pi G\dot\phi^2.
\label{friedmann1}
\end{eqnarray}

If the potential satisfies the so-called slow-roll conditions,
\beq
\label{epsilon_eta}
\epsilon_V\equiv {\mP^2\over 2}\left({V'\over V}\right)^2 \ll 1, \quad \eta_V\equiv \mP^2 {V''\over V}\ll 1,
\eeq
where 
$m_P\equiv (8\pi G)^{-1/2}$ is the reduced Planck mass, the evolution can enter into a  {\it slow-roll inflationary regime}
where the kinetic energy of the scalar field  in (\ref{friedmann1}) 
and the acceleration $\ddot\phi$ in the  Klein-Gordon equation 
(\ref{KG}) can be neglected. In this regime, the equations governing the evolution of the scale factor and the scalar field reduce to 
\beq
H^2\simeq{8\pi G\over 3} V, \qquad 3H\dot \phi+V'\simeq 0.
\eeq
A useful quantity, which can then be easily derived, is the number of e-folds $N\equiv \ln (a_{\rm end}/a)$ between some instant during inflation and 
the end of inflation (or, more precisely, the end of the slow-roll regime):
\beq
N(\phi)\simeq \int_\phi^{\phi_{end}}{V\over m_P^2 V'}d\phi
\eeq
For any model of inflation, the number of e-folds between the onset of inflation and reheating must be sufficient, typically of the order of $60$, in order to solve the horizon problem discussed earlier. In order to get more 
detailed constraints on the models from observations, it is necessary to go beyond the homogeneous description and consider cosmological perturbations.

\subsection{Cosmological perturbations}
In the theory of linear cosmological perturbations\footnote{See \cite{cargese} for a basic presentation and \cite{mfb} for a
detailed review.}, both the matter (i.e. the scalar field for inflation) and the geometry, i.e. the metric, are perturbed. Restricting ourselves to scalar perturbations, the metric can be written as
\beq 
ds^2= -(1+2A)dt^2+ 2 a(t) \partial_iB\, dx^idt+
a^2(t)\left[(1-2\psi)\delta_{ij}
+2\partial_i\partial_jE\right]dx^idx^j,
\eeq
where $\psi$ is directly related to the intrinsic curvature of constant time hypersurfaces, according to the relation
\beq
{}^{(3)}R=\frac{4}{a^2}\partial^2\psi.
\eeq
The metric perturbations are modified in a change of coordinates. It is thus useful (although not necessary) to define gauge-invariant quantities, such as the curvature perturbation on uniform energy hypersurfaces, defined by
\beq
\label{zeta}
-\zeta\equiv \psi+\frac{{ H}}{\dot\rho}\delta\rho
=\psi-\frac{\delta\rho}{3(\rho+p)},
\eeq
or 
the comoving curvature perturbation, 
\beq
{\cal R}\equiv \psi-\frac{H}{\rho+p}\delta q \,, 
\eeq
where $\delta q$ is the scalar part of the momentum density ($\delta T_i^0\equiv \partial_i\delta q$). 
Using the linearized Einstein's equations, it can be shown that these two quantities are related via
\beq
\zeta=-{\cal R}-{2\rho\over 3(\rho+P)}\left({k\over aH}\right)^2\Psi
\eeq
where
\beq
\label{Psi}
\Psi = \psi + a^2 H (\dot{E}-B/a).
\eeq
The quantity $\zeta$ is particularly interesting because it is conserved on large scales when the matter perturbations are adiabatic, i.e. when they satisfy 
\beq
\delta P_{\rm nad}\equiv 
\delta p-{{\dot p}\over {\dot\rho}}\, \delta\rho=0.
\eeq
This property, which is well-known for linear perturbations, can be seen as the consequence of a more general 
result. Indeed, the conservation of the energy-momentum tensor 
for any perfect fluid, characterized by the energy density $\rho$, the pressure $p$ and the four-velocity $u^a$, leads 
to the {\it exact} relation\cite{Langlois:2005ii,Langlois:2005qp}
\beq
\label{dot_zeta}
\dot\zeta_a\equiv {\cal L}_u\zeta_a=
-{\Theta\over{3(\rho+p)}}\left( \nabla_a p -
\frac{\dot p}{\dot \rho} \nabla_a\rho\right), 
\eeq
where we have defined 
\beq
 \zeta_a\equiv 
\nabla_a\alpha-\frac{\dot\alpha}{\dot\rho}\nabla_a\rho, \quad \Theta=\nabla_a u^a, \quad \alpha=\frac{1}{3}\int d\tau \,
\Theta,
\eeq
and where a dot on scalar quantities denotes here a derivative along $u^a$ (e.g. $\dot\rho\equiv u^a\nabla_a\rho$). This identity is valid for any spacetime geometry and does not rely on Einstein's equations. In the cosmological context, $\alpha$ can be interpreted as a non-linear generalization, according to an observer following the fluid, of the number of e-folds of the scale factor. Introducing an explicit coordinate system and linearizing (\ref{dot_zeta}) leads to the familiar result of the linear theory.

During inflation, it is easier to work with the perturbation $\R$, since in this case
\beq
{\cal R}=\psi+\frac{H}{\dot\phi} \delta\phi\, .
\eeq
Because of the constraints arising from Einstein's equations, the scalar metric perturbations and the scalar field perturbation are not independent. In fact, there is only one degree of freedom which can be expressed in terms of the combination
\beq
v=a\left(\delta\phi+\frac{\dot\phi}{H} \psi\right)\equiv a \, Q\, ,
\eeq
where $Q$ represents the scalar field perturbation in the spatially flat gauge (where $\psi=0$).
The quadratic action governing the dynamics of this degree of freedom can be obtained from the expansion up to second order of the full action. One finds (see e.g. \cite{mfb})
\beq
S_v={1\over 2}\int d\tau  \, \, d^3x\, \left[{v'}^2+\partial_i v \partial ^i v
+{z''\over z}v^2\right],
\eeq
where a prime denotes a derivative with respect to the conformal time $\tau=\int dt/a(t)$, and 
with
\beq
\label{z}
z=a\frac{\dot\phi}{H}.
\eeq
To quantize this system, one considers $v$ as a quantum field and one 
 decomposes it  as 
\beq
\label{Fourier_quantum}
\hat v (\tau, \vec x)={1\over (2\pi)^{3/2}}\int d^3k \left\{{\hat a}_{\vec k} v_{k}(\tau) e^{i \vec k.\vec x}
+ {\hat a}_{\vec k}^\dagger v_{k}^*(\tau) e^{-i \vec k.\vec x} \right\},
\eeq
where  the $\hat a^\dagger$ and  $\hat a$ are 
 creation and annihilation operators , which satisfy the 
usual commutation rules 
\beq
\label{a}
\left[ {\hat a}_{\vec k}, {\hat a^\dagger}_{\vec k'}\right]= \delta(\vec k-\vec k')\, ,
\quad
\left[ {\hat a}_{\vec k}, {\hat a}_{\vec k'}\right]= 
\left[ {\hat a^\dagger}_{\vec k}, {\hat a^\dagger}_{\vec k'}\right]= 0\, .
\eeq
The action  implies that  the conjugate momenta for $v$ is
$v'$. Therefore,  the canonical quantization  
for $\hat v$ and its conjugate momentum leads to  the condition
\beq
v_{k} {v'_{k}}^*-v_{k}^*v'_{k}=i\,.
\label{wronskien}
\eeq
The complex function  $v_{k}(\tau)$ satisfies the equation of motion 
\beq
v''+\left( k^2-\frac{z''}{z}\right) v=0.
\eeq
In the slow-roll limit,  $z''/z\simeq 2/\tau^2$, and one can use the solution for a de Sitter spacetime (where $H$ is constant). Note that this is only an approximation as the Hubble parameter is decreasing with time, but a very good one,  when the slow-roll parameters are small, during the short time when the scale of interest crosses out the Hubble 
radius ($k\sim a H$).
 Requiring that the solution on small scales behaves like the Minkowski vacuum selects  the particular solution 
\beq
v_{k}\simeq  \frac{1}{\sqrt{2k}}e^{-ik \tau }\left(1-{i\over k \tau}\right),
\eeq
where the normalization is imposed by the condition (\ref{wronskien}). 
This implies that the power spectrum of the scalar field fluctuations is given by
\beq
\label{spectrum_scalar}
{\cal P}_{Q}=\frac{k^3}{2\pi^2}|v_k|^2\frac{1}{a^2}\simeq\frac{H^2}{4\pi^2},
\eeq
where the quantities on the right hand side are evaluated at 
 {\it Hubble crossing}. This can be translated into the power spectrum of the curvature perturbation 
 ${\cal R}$, by noting that ${\cal R}=aQ/z$. One thus gets
\beq
\label{spectrum_inflation}
{\cal P}_{\cal R}=\frac{k^3}{2\pi^2}\frac{|v_{k}|^2}{z^2}\simeq\left(\frac{H^4}{4\pi^2 \dot\phi^2}\right)_{|k=aH}=
\frac{1}{2\mP^2\epsilon_*}\left(\frac{H_*}{2\pi }\right)^2 \;,
 \eeq
where $\epsilon_*$ is the slow-roll
parameter defined in (\ref{epsilon_eta}), the label $*$ denoting its value at Hubble crossing.

In single-field inflation, since ${\cal R}$ is conserved on large scales (as ${\cal R}$ and $\zeta$ coincide on large scales), the above expression, evaluated at Hubble crossing, determines the amplitude 
of the curvature perturbation just before the modes reenter the Hubble radius and thus sets 
the initial conditions for cosmological perturbations. 
As we will see in the next section, this is no longer true for inflationary models involving several scalar fields.

We have focused so far  our attention on scalar perturbations, which 
are the most important in cosmology. Tensor perturbations, or primordial 
gravitational waves, if ever detected in the future, would be a remarkable 
probe of the early universe. In the inflationary scenario, 
like scalar perturbations, primordial 
gravitational waves are generated from vacuum quantum fluctuations.
 Let us now explain briefly this mechanism.

Starting from the metric with tensor perturbations,
\beq
ds^2= a^2(\tau)\left[-d\tau^2+\left(\delta_{ij}
+\bar{E}_{ij}\right) dx^idx^j\right],
\eeq
where $\bar{E}_{ij}$ is transverse traceless (i.e. $\partial^i\bar{E}_{ij}=0$ and $\delta^{ij}\bar{E}_{ij}=0$),
the action expanded at  second order in the perturbations yields 
\beq
S^{(2)}_g={1\over 64\pi G}\int  d\tau\,  d^3x \, a^2 
\eta^{\mu\nu}\partial_\mu \bar E^i_j\partial_\nu\bar E^j_i,
\eeq
where $\eta^{\mu\nu}$ denotes the Minkoswki metric.
Apart from the tensorial nature of $E^i_j$, this action is quite 
similar to that of a massless scalar field in a FLRW universe, 
up to a renormalization 
factor $1/\sqrt{32\pi G}$.
The decomposition
\beq
a\bar E^i_j=\sum_{\lambda=+,\times}
\int {d^3k\over (2\pi)^{3/2}} 
 v_{k,\lambda}(\tau)\epsilon^i_j({\vec k};\lambda) e^{i \vec k.\vec x}
\eeq
where the $\epsilon^i_j({\vec k};\lambda)$ are the polarization tensors,
shows that the gravitational waves are essentially equivalent to two 
massless
scalar fields (for each polarization) $\phi_\lambda=m_P\bar E_\lambda /2$.

The total power spectrum is thus immediately deduced from (\ref{spectrum_scalar}): 
\beq
{\cal P}_T=2\times {4\over \mP^2}\times \left({H\over 2\pi}\right)^2, 
\eeq
where the first factor comes from the two polarizations, the 
second from the renormalization with respect to a canonical scalar field, 
the last term being the power spectrum for a scalar field derived earlier.
In summary, the tensor power spectrum is
\beq
{\cal P}_T={2\over \pi^2}\left({H_*\over \mP}\right)^2,
\label{spectrum_tensor}
\eeq
where the label $*$ recalls that the Hubble parameter, which can be slowly evolving
during inflation, must be evaluated when the relevant scale crossed out the 
Hubble radius.

\section{Multi-field inflation}
So far, the simplest models of inflation are compatible with the data but it is instructive to study more refined models for at least two reasons. First, because models inspired by high energy physics are usually more complicated than the simplest phenomenological inflationary models. Second, because these generalized models will  give us an idea of how much the future data will be able to pin down some specific region in the ``space'' of models. 

In this section, we first discuss some potentially  observational signatures of more sophisticated models, namely entropic perturbations and non-Gaussianities, 
which, if ever detected, would provide invaluable additional clues on the early Universe. We then turn to some specific scenarios: 
the curvaton mechanism, and multi-field inflation with non standard kinetic terms illustrated by 
multi-field Dirac-Born-Infeld inflation. 

\subsection{Adiabatic and entropic perturbations}
Before considering various types of multi-field scenarios, it is instructive to discuss  potentially observational 
effects that would discriminate between multi-field and single-field inflation. In the case of single field inflation, all perturbations of the cosmological fluid, which consists of photons, neutrinos, baryons and cold dark matter (CDM) particles 
ultimately originate from the primordial scalar field fluctuations and satisfy the adiabaticity property, $\delta(n_m/n_r)=0$, or 
\beq
\frac{\delta\rho_m}{\rho_m}=\frac{3\delta\rho_r}{4\rho_r},
\eeq
where the index $m$ stands for a non-relativistic species (either baryonic matter or CDM) and 
$r$ for a relativistic species (photons or neutrinos). 

By contrast, in a multi-field scenario, one can envisage a richer spectrum of possibilities, such as the existence of 
non-adiabatic, or entropic perturbations, for example between CDM and photons, defined by
\beq
S\equiv 
\frac{\delta \rho_c}{\rho_c}
-{3\over 4}\frac{\d\rho_\gamma}{\rho_\gamma}.
\eeq 
Interestingly, this entropic perturbation could be correlated with the adiabatic perturbation \cite{Langlois:1999dw,Langlois:2000ar}.
The adiabatic and entropic perturbations lead to a different peak structure  in the CMB fluctuations and, therefore, CMB measurements can potentially distinguish between these two types of perturbations. On large angular scales, 
one can show for instance that \cite{Langlois:1999dw}
\beq
\frac{\delta T}{T}\simeq \frac{1}{5}
\left({\cal R}
-2 {\cal S}\right).
\eeq
The combined impact of adiabatic and entropic perturbations crucially depends on their correlation
\beq
\beta=
\frac{ {\cal P}_{_{{\cal S},{\cal R}}}}
{\sqrt{{\cal P}_{_{{\cal S}}}{\cal
  P}_{_{\cal R}}}}.
\eeq
Parametrizing the relative amplitude between the two types of perturbations  by a coefficient $\alpha$, 
\beq
\frac{{\cal P}_{{\cal S}}}{{\cal
  P}_{\cal R}}\equiv \frac{\alpha}{1-\alpha}\, ,
\eeq
the present constraints  on the entropy contribution are $\alpha_0<0.067 \ (95\%\, {\rm C.L.})$ in the 
uncorrelated case ($\beta=0$) and $\alpha_{-1}<0.0037 \ (95\%\, {\rm C.L.})$ in the totally anti-correlated case 
($\beta=-1$) \cite{wmap5}. 

\subsection{Non-Gaussianities}
Another interesting feature of some early Universe models is to produce primordial perturbations with a significant non-Gaussianity, which could be detected in future observations (see \cite{Bartolo:2004if} for a review). 
Note that, in contrast with entropic perturbations, a significant 
non-Gaussianity is not specific to multi-field models as single field models with non-standard kinetic terms can produce a (relatively) high level of non-Gaussianity. 

The most natural estimate of non-Gaussianity is the bispectrum defined, in Fourier space, by 
\beq
\langle
 \zeta_{\bk_1} \zeta_{\bk_2} \zeta_{\bk_3} 
\rangle \equiv (2 \pi)^3
\delta^{(3)}(\sum_i \bk_i) 
B_\zeta (k_1,k_2,k_3).
\eeq
Equivalently, one often uses the so-called $f_{NL}$ parameter, which can be defined in general by
\beq
\frac{6}{5}
f_{\rm NL} 
\equiv \frac{\Pi_i k_i^3}{
\sum_i k_i^3}
\frac{B_{\zeta}}{4 \pi^4 \mP^2\PP_\zeta^2}.
\eeq

In the context of multi-field inflation, the
so-called $\delta N$-formalism \cite{deltaN} is particularly
useful to evaluate the primordial non-Gaussianity generated on large scales
\cite{Lyth:2005fi}. The idea is to describe, on scales larger than the Hubble radius, 
the non-linear evolution of perturbations generated during  inflation in
terms of the perturbed expansion from an initial hypersurface
(usually
taken at Hubble crossing
during inflation) up to a final uniform-density
hypersurface (usually during the radiation-dominated era).
Using the expansion
\beq
\zeta \simeq  \sum_I N_{,I} \delta \varphi_*^I + \frac{1}{2}
\sum_{IJ} N_{,IJ} \delta \varphi_*^I \delta \varphi_*^J
\eeq
yields the expression
\begin{eqnarray}
\langle \zeta_{\bk_1} \zeta_{\bk_2} \zeta_{\bk_3} \rangle &=&
\sum_{IJK} N_{,I} N_{,J} N_{,K} \langle \delta \varphi^I_{\bk_1}
\delta \varphi^J_{\bk_2} \delta \varphi^K_{\bk_3}\rangle + \nonumber
\\ &&  \frac{1}{2} \sum_{IJKL} N_{,I} N_{,J} N_{,KL} \langle \delta
\varphi^I_{\bk_1} \delta \varphi^J_{\bk_2} (\delta \varphi^K \star
\delta \varphi^L)_{\bk_3}\rangle
+{\rm perms}. \nonumber \\ &&
\label{tpf}
\end{eqnarray}
If the scalar field fluctuations are quasi-Gaussian, one can ignore their three-point correlations and, after substituting 
\beq
\langle
\delta\varphi^I_{\bk_1} \delta\varphi^J_{\bk_2} \rangle = (2
\pi)^3 \delta_{IJ} \delta^{(3)} (\bk_1 +
\bk_2)  \frac{2 \pi^2}{k_1^3} \PP_* (k_1) , \qquad \PP_*(k)  \equiv \frac{H_*^2}{4
\pi^2},
\eeq
one gets
\beq
\label{f_local}
\frac{6}{5}f_{\rm NL} = 
  \frac{
N_{I} N_{J} N^{IJ}}{( N_{K}N^K)^2}\, .
\eeq
The present observational constraints \cite{wmap5} are $ -9< f_{NL}^{\rm (local)}<111 \quad (95\% \, {\rm CL})$ and 
$-151< f_{NL}^{\rm (equil)}<253 \quad (95\%\,  {\rm CL})$, for respectively, the local non-linear coupling parameter
and the equilateral non-linear coupling parameter (characterizing the amplitude of the bispectrum of the equilateral 
configurations in which the three wave vectors forming a triangle in Fourier space have the same length).

\subsection{The curvaton scenario}
\def\curv{\chi}
\def\zetar{\zeta_{\rm r}}
\def\zetarad{\zeta_{\rm r}}
\def\i{{\rm inf}}

 The curvaton (see \cite{curvaton}) is a weakly coupled scalar field,
$\curv$,
which is light relative to the Hubble
rate during inflation, and hence acquires an almost
scale-invariant spectrum and effectively Gaussian distribution of
perturbations, $\delta\curv$, during inflation, 
\beq
\label{spectrum_curvaton}
{\cal P}_{\delta\chi}=\left(\frac{H}{2\pi}\right)^2.
\eeq

 After inflation
the Hubble rate drops and eventually the curvaton becomes
non-relativistic so that its energy density grows relative to
radiation, until it contributes a significant fraction of the
total energy density, $\Omega_\curv\equiv\bar\rho_\curv/\bar\rho$,
before it decays. Hence the initial curvaton field perturbations
on large scales can give rise to a primordial density perturbation
after it decays.

The non-relativistic curvaton (mass $m\gg H$), before it decays,
can be described by a pressureless, non-interacting fluid with
energy density
 \begin{equation}
\rho_\curv = m^2 \curv^2 \;,
 \end{equation}
where $\curv$ is the rms amplitude of the curvaton field, which
oscillates on a timescale $m^{-1}$ much less than the Hubble time
$H^{-1}$.
The corresponding perturbations are characterized, using (\ref{zeta}) and (\ref{spectrum_curvaton}),
\beq
\zeta_\chi=\left(\frac{\delta\rho_\chi}{3\rho_\chi}\right)_{\rm flat}
\quad \Rightarrow {\cal P}_{\zeta_\chi}\simeq \frac{H^2}{9\pi^2 \chi^2}.
\eeq
When the curvaton decays into radiation, its perturbations are converted into perturbations of the resulting radiation
fluid. The subsequent perturbation is described by
 \begin{equation}
 \zetar = r \zeta_\curv +(1-r)\zeta_\i\,, \quad r \equiv \frac{3\Omega_{\curv,{\rm decay}}}{4-\Omega_{\curv,{\rm
decay}}}\, .
 \end{equation}
This implies that
the power spectrum for the primordial adiabatic perturbation
$\zetarad$ can be expressed as
 \beq
 \label{finalzetar}
  {\cal P}_{\zetarad}={\cal P}_{\zeta_\i}+r^2{\cal P}_{\zeta_\chi} \;.
 \eeq
where ${\cal P}_{\zeta_\i}$ is given by (\ref{spectrum_inflation})
 in the case of standard single field inflation. In most cases, the 
 inflaton contribution is supposed to be negligible but one can also envisage  mixed
 inflaton-curvaton scenarios where both contribute (see e.g. \cite{Langlois:2004nn}).  

Interestingly, the curvaton scenario can give rise to a significant non-Gaussianity of the local type, 
since the expression (\ref{f_local}) yields \cite{Lyth:2005fi}
\beq
f_{NL}^{\rm local}= \frac{5}{4r}-\frac{5}{3}-\frac{5}{6}r\, .
\eeq
The curvaton can also produce some isocurvature perturbations \cite{Lyth:2002my}, possibly with 
a significant non-Gaussianity \cite{lvw08}.

\subsection{Multi-inflaton scenario}
We now consider  multi-field models,  which can be described by an
action of the form
\beq
\label{P}
S =  \int d^4 x \sqrt{-g}\left[\frac{R}{16\pi G}   +   P(X^{IJ},\phi^K)\right] 
\eeq
where $P$ is an arbitrary function of $N$ scalar fields and of the  kinetic term
\beq
\label{X}
X^{IJ}=-\half \nabla_\mu \p^I  \nabla^\mu \p^J.
\eeq
The very general form (\ref{P})  can be seen as an extension of k-inflation 
\cite{ArmendarizPicon:1999rj} to the case of  several scalar fields. 

A more restrictive class of models, considered in \cite{lr08}, consists of Lagrangians that depend on a global kinetic term 
$X=G_{IJ}X^{IJ}$ where $G_{IJ} \equiv G_{IJ}(\p^K)$ is an arbitrary metric on the $N$-dimensional field space.
By defining $P=X-V$, one recovers in particular multi-field models with an action of the form
\beq
S=\int d^4x\sqrt{-g}\left(-{1\over 2}\, 
G_{IJ}(\phi)
\, \partial^\mu\phi^I\partial_\mu\phi^J
-V(\phi)\right),
\eeq
where  a flat metric in field space ($G_{IJ}=\delta_{IJ}$) corresponds to standard kinetic terms.

The relations obtained in the previous section for  the single field model can then be generalized.
The energy-momentum tensor, derived from (\ref{P}),   is of the form
\beq
T^{\mu \nu} = P g^{\mu \nu} + P_{<IJ>}  \partial^\mu \phi^I \partial^\nu \phi^J\,,
\label{Tmunu}
\eeq
where $P_{<IJ>}$ denotes the partial derivative of $P$ with respect to $X^{IJ}$ (symmetrized with respect to the indices $I$ and $J$).
The equations of motion for the scalar fields, which can be seen as generalized Klein-Gordon equations, are obtained from the variation of the action with respect to $\phi^I$. One finds
\beq
\nabla_{\mu} \left(  P_{<IJ>} \nabla^\mu \phi^J \right) + P_{,I} = 0\,.
\label{KG-general}
\eeq
where $P_{,I}$ denotes the partial derivative of $P$ with respect to $\phi^I$.

In a spatially flat FLRW  spacetime, with metric
\beq
ds^2=-dt^2+a^2(t)d{\vec x}^2,
\eeq
the scalar fields are homogeneous, so that $X^{IJ}=\dot\phi^I\dot\phi^J/2$, and the energy-momentum tensor reduces to that of a perfect fluid
with energy density 
\beq
\rho=2  P_{<IJ>}  X^{IJ} - P\,,
\label{rho}
\eeq
and pressure $P$. 
 The evolution of the scale factor $a(t)$ is governed by the Friedmann equations, which can be written in the form
\beq
H^2 = \frac{1}{3} \left(2  P_{<IJ>}  X^{IJ} - P \right)\, , \qquad
\dot{H} = - X^{IJ} P_{<IJ>} \,.
\label{Friedmann2}
\eeq
The equations of motion for the scalar fields reduce to
\beq
\left( P_{<IJ>} + P_{<IL>,<JK>} \dot{\phi}^L \dot{\phi}^K  \right) \ddot{\phi}^J+ 
 \left( 3 H P_{<IJ>} + {P}_{<IJ>,K} \dot{\phi}^K\right)\dot{\phi}^J - P_{,I} = 0\, ,
\label{KG1}
\eeq
where $P_{<IL>,<JK>}$ denotes the (symmetrized) second derivative of $P$ with respect to $X^{IL}$ and $X^{JK}$.

The expansion up to second order in the linear perturbations of the action (\ref{P}) is useful to obtain the classical equations of motion for the perturbations and to calculate the spectra of the primordial perturbations generated  during inflation, as we have seen in the previous section for a single scalar field.  
 Working for convenience with the scalar field perturbations 
$Q^I$ defined in the spatially flat gauge, the
 second order action can be  written in the rather simple form 
 \begin{eqnarray}
S_{(2)} &=& \frac{1}{2} \int {\rm d}t \, {\rm d}^3x \, a^3 \left[ 
\left(P_{<IJ>} + 2 P_{<MJ>,<IK>}X^{MK}\right) \dot{Q}^{I}\dot{Q}^{J}  
\right.
\cr
&& 
\qquad \left.
- P_{<IJ>} h^{ij} \partial_iQ^I \partial_jQ^J 
 - {\cal M}_{KL}Q^K Q^L 
 + 2 \,\Omega_{KI}Q^K \dot{Q}^I  \right] \qquad
 \label{S2}
\end{eqnarray}
where the explicit expressions for the mass matrix 
${\cal M}_{KL}$
and for the mixing matrix 
$\Omega_{KI}$
can be found in \cite{lrst08b}. 

\subsection{Example: multi-field DBI inflation}
Recent years have seen an intensive effort  to construct models of inflation within string theory (for a recent review, see e.g.~\cite{McAllister:2007bg}).
Effective descriptions of string theory at low energies are based on 10-dimensional spacetimes, which can be 
related to our apparent 4-dimensional spacetime by assuming that six dimensions span a 6-dimensional compact internal
manifold. In addition to the fundamental strings, string theories contain higher dimensional objects called D-branes (where the D stands for Dirichlet boundary conditions). An interesting suggestion was to identify the inflaton(s) with 
the position of some D-brane in the internal compact space. Effective four-dimensional inflation could thus result
from the motion of a D-brane in the internal dimensions. This type of scenario is called {\it brane inflation}
 \cite{Dvali:1998pa}. 

Let us consider, for instance, a D3-brane with tension $T_3$ evolving in a 10-dimensional geometry described by the metric
\beq
ds^2 = h^{-1/2}(y^K)\,g_{\mu \nu}dx^\mu dx^\nu + h^{1/2}(y^K)\, G_{IJ}(y^K)\, dy^I dy^J \equiv H_{AB} dY^A dY^B
\eeq
with coordinates $Y^A=\left\{x^\mu, y^I\right\}$, 
where $\mu=0,\ldots 3$ and $I=1,\ldots, 6$.  

The motion of the brane is described by a Dirac-Born-Infeld Lagrangian, 
\beq
\label{L}
L =- T_3\sqrt{-\det{\gamma_{\mu \nu}}}
\eeq
which depends on the determinant of the induced metric on the 3-brane,
\beq
\gamma_{\mu \nu} =H_{AB} \partial_\mu Y_{\rm (b)}^A \partial_\nu Y_{\rm (b)}^B =
 h^{-1/2} \left( g_{\mu \nu}  + h \, G_{IJ} \partial_\mu \varphi^I \partial_\nu \varphi^J \right)
\label{DBIzero}\, ,
\eeq
where  the functions  $Y_{\rm (b)}^A(x^\mu)=(x^\mu, \varphi^I(x^{\mu}))$ define the  brane embedding 
 (with the $x^\mu$ being the spacetime coordinates on the brane). 
After various rescalings, one ends up with a Lagrangian of the form
\beq
P= -\frac{1}{f(\bfphi^I)}\left(\sqrt{{\cal D}}-1\right) -V(\bfphi^I)
\label{DD}
\eeq
with
\beq
{\cal D}\equiv  \det(\delta^{\mu}_{\nu}+f \, G_{IJ}\partial^{\mu} \p^I \partial_{\nu} \p^J )\,,
\label{Ddef}
\eeq
and where we have also included  potential terms, which arise from the brane's interactions with bulk fields or other branes. 

An interesting situation is when the brane moves in a higher dimensional warped conical geometry, along the 
radial direction. If one ignores the angular internal coordinates,   the four-dimensional effective action reduces to
\beq
S=\int d^4x \sqrt{-g}\left[- \frac{1}{f}\left(\sqrt{1+f  \, \partial_\mu\phi\partial^\mu \phi} -1\right)
-V(\phi)\right]\, ,
\eeq
which
depends on a single scalar field but with non standard kinetic terms.
This action belongs to the class of \textit{k}-inflation models
\cite{ArmendarizPicon:1999rj} characterized by a Lagrangian of the form $P(X,\phi)$, where 
$X=-\partial_\mu\phi \partial^\mu\phi/2$.
If $f\dot\phi^2\ll 1$, one can expand the square root in the Lagrangian and one recovers the usual kinetic term familiar to slow-roll inflation. But there is another regime, called DBI inflation \cite{dbi}, corresponding
to the ``relativistic'' limit
\beq
1-f  \, \dot\phi^2\ll 1 \Rightarrow |\dot\phi|\simeq 1/\sqrt{f}\, ,
\eeq
which does not require a very flat potential as in standard slow-roll inflation.

Allowing the brane to move in the angular directions  leads to a multi-field scenario, since each brane coordinate in the extra dimensions gives rise to a scalar field from the effective four-dimensional point of view. 
The corresponding  Lagrangian (\ref{DD}) can be written in the generic form (\ref{P}) that depends on the kinetic terms
$X^{IJ}$ defined in (\ref{X}) by noting that  \cite{lrst08a}
\begin{eqnarray}
\label{def_explicit}
{\cal D}&=&1-2f G_{IJ}X^{IJ}+4f^2 X^{[I}_IX_J^{J]} -8f^3 X^{[I}_IX_J^{J} X_K^{K]}+16f^4 X^{[I}_IX_J^{J} X_K^{K}X_L^{L]}
\cr
&\equiv & 1-2f \tX\, \, 
\end{eqnarray}
where the field indices are lowered by the field metric $G_{IJ}$.

The dynamics of the linear perturbations can be obtained from the general expressions (\ref{S2}). 
Alternatively, one can use the results of \cite{lr08} for Lagrangians of the form $P=P(X,\phi^K)$, where $X=G_{IJ}X^{IJ}$, by writing the Lagrangian (\ref{DD}) as a function of $\tX$, introduced in (\ref{def_explicit}), 
so that  
$P(X^{IJ},\phi^K) =  \tilde{P}(\tilde{X},\phi^K)$ (note that $\tX$ and $X$ coincide in the homogeneous background). 
What characterizes the DBI multi-field Lagrangian is that all linear perturbations propagate with a common velocity,
namely the effective speed of sound defined by 
\beq
c_s=\sqrt{1-2fX}\, .
\eeq

For simplicity, let us now concentrate on a two-field scenario.  It is then useful to 
decompose the scalar field perturbations into adiabatic and entropic modes \cite{Gordon:2000hv}, namely
\beq
\label{decomposition}
Q^I=Q_\s e_\sigma^I+Q_s e^I_s\,,
\eeq
where 
\beq
\label{e1}
e_\sigma^I=\frac{\dot{\p^I}}{\sqrt{2X}},
\eeq
is the unit vector along the inflationary trajectory in field space and 
 the entropy vector $e^I_s$ is the  unit vector orthogonal to the adiabatic vector $e_\sigma^I$, i.e.
\beq
G_{IJ}e_s^I e_s^J=1, \qquad G_{IJ}e_s^I e_\sigma^J=0.
\eeq
As in standard inflation discussed in the previous section, it is more convenient, after going to conformal time $\tau = \int {dt}/{a(t)}$, to work in terms of the canonically normalized fields
\beq
v_{\s}=\frac{a}{c_s^{3/2}} \, Q_{\s} \,,\qquad \,v_{s}=\frac{a}{\sqrt{c_s}}\, Q_s\,,
\label{v}
\eeq
which lead to the second order action
\begin{eqnarray}
\label{S_v}
S_{(2)}&=&\frac{1}{2}\int {\rm d}\tau\,  {\rm d}^3x \Big\{ 
  v_\s^{\prime\, 2}+ v_s^{\prime\, 2} -2\xi v_\s^\prime v_s-c_s^2 \left[(\partial v_\s)^2 + (\partial v_s)^2\right] 
\cr
&& 
\left.
\qquad
+\frac{z''}{z} v_\s^2
+\left(\frac{\alpha''}{\alpha}-a^2 \mu_s^2\right) v_s^2+2\, \frac{z'}{z}\xi v_\s v_s\right\}
\end{eqnarray}
with
\beq
\xi=\frac{a}{\dot \s \tP_{,X} c_s}[(1+c_s^2)\tP_{,s}-c_s^2 \dot\s^2 \tP_{,Xs}]\,, \qquad \dot\s\equiv\sqrt{2X}\, ,
\label{11}
\eeq
and
where we have introduced the two background-dependent  functions 
$z=a \dot \s/({H c_s^{3/2}})$ and $\alpha={a}/{\sqrt{c_s}}$.

The equations of motion derived from the action (\ref{S_v}) can  be written in the compact form
\begin{eqnarray}
v_{\s}''-\xi v_{s}'+\left(k^2c_s^2 -\frac{z''}{z}\right) v_{\s} -\frac{(z \xi)'}{z}v_{s}&=&0\,.
\label{eq_v_sigma}
\\
v_{s}''+\xi  v_{\s}'+\left(k^2 c_s^2- \frac{\alpha''}{\alpha}+a^2\mu_s^2\right) v_{s} - \frac{z'}{z} \xi v_{\s}&=&0\,.
\label{eq_v_s}
\end{eqnarray}
Assuming that  the coupling $\xi$ is very small and that  $|\mu_s^2|/H^2\ll 1$ when the scales of interest cross out the sound horizon, i.e. $kc_s=aH$, the above system leads to the amplification of the vacuum fluctuations 
at sound horizon crossing for both adiabatic and entropic degrees of freedom. 

Following the standard procedure outlined in the previous section, 
  one selects the  positive frequency solutions of Eqs.~(\ref{eq_v_sigma}) and (\ref{eq_v_s}), which  correspond to the usual vacuum on very small scales:
  \beq
v_{\s\, k} \simeq v_{s\, k} \simeq  \frac{1}{\sqrt{2k c_s}}e^{-ik c_s \tau }\left(1-\frac{i}{k c_s\tau}\right)\, .
\eeq
As a consequence, the power spectra for $v_\s$ and $v_s$ after sound horizon crossing  have the same amplitude\beq
{\cal P}_{v_\s}={\cal P}_{v_s}=\frac{k^3}{2\pi^2}|v_{\s\, k}|^2\simeq\frac{H^2 a^2}{4\pi^2 c_s^3}.
\eeq
 However, in terms of the initial fields $Q_\s$ and $Q_s$, one finds, using (\ref{v}), 
\beq
\label{power_sigma}
{\cal P}_{Q_\s*}\simeq\frac{H^2}{4\pi^2 }, \quad {\cal P}_{Q_s*}\simeq\frac{H^2}{4\pi^2 c_s^2},
\eeq
(the subscript $*$ indicates that the corresponding quantity is evaluated at sound horizon crossing $k c_s=aH$)
which shows that, for small $c_s$, the entropic modes are {\it amplified} with respect to the adiabatic modes:
\beq
Q_{s*}\simeq \frac{Q_{\sigma*}}{c_s}.
\eeq

In order to confront the predictions of inflationary models to cosmological observations, it is useful to rewrite the scalar field perturbations in terms of geometrical quantities.
Using the relation
 \beq
{\cal R}=\frac{H}{\dot \s}Q_{\s}\,,
\label{R}
\eeq
one  recovers the usual {\it single-field} result \cite{Garriga:1999vw} that the power spectrum for $\R$ at sound horizon crossing is given by
\beq
{\cal P}_{\cal R_*}=\frac{k^3}{2\pi^2}\frac{|v_{\s\, k}|^2}{z^2}\simeq\frac{H^4}{4\pi^2  \dot\s^2 }=\frac{H^2}{8\pi^2 \epsilon c_s }\,,
\label{power-spectrum-R}
\eeq
where $\epsilon=-\dot H / H^2\,$.

However, in contrast with single field inflation, the curvature perturbation can be subsequently modified if there 
is a transfer between entropy and adiabatic modes \cite{sy} (see also \cite{Lalak:2007vi} for a recent numerical 
treatment). This transfer from the entropic to the adiabatic modes  can be parametrized by the transfer coefficient  which appears in the formal solution $\R=\R_*+T_{ {\cal R}  {\cal S} } \cal S_*$ of the evolution equations. For convenience, we use the entropy perturbation, which we denote ${\cal S}$, whose power spectrum
at sound horizon crossing is the same as that of the curvature perturbation, i.e.
\beq
{\cal S}=c_s\frac{H}{\dot \s}Q_{s}\, ,
\label{S}
\eeq
so that ${\cal P}_{\cal S_*}={\cal P}_{\cal R_*}$.
The final curvature power-spectrum is thus given by
\beq
{\cal P}_{\cal R}=(1+T_{{\cal R} {\cal S}}^2) {\cal P}_{\cal R_{*}}=\frac{{\cal P}_{\cal R_{*}}}{{\rm cos^2} \Theta}\, ,
\label{observed-spectrum}
\eeq
where we have introduced  the ``transfer angle'' $\Theta$ ($\Theta=0$ if there is no transfer and $|\Theta|=\pi/2$ if the final curvature perturbation is mostly of entropic origin) by
\beq
{\sin} \Theta =\frac{T_{ {\cal R}  {\cal S} }}{\sqrt{1+T^2_{ {\cal R}  {\cal S} }}}\,.
\label{correlation-result}
\eeq

The power spectrum for the tensor modes is still governed by the transition at {\it Hubble radius} and its amplitude, given by (\ref{spectrum_tensor}),
is much smaller than the curvature amplitude in the small $c_s$ limit. The tensor to scalar ratio is 
\beq
r \equiv \frac{{\cal P}_{\cal T}}{{\cal P}_{\cal R}}=16\, \epsilon\, c_s {\cos^2} \Theta.
\label{r}
\eeq
Interestingly this expression combines the result of $k$-inflation, where the ratio is suppressed by the sound speed $c_s$, and that of standard multi-field inflation\cite{Bartolo:2001rt}.

It is also possible to compute the non-Gaussianities generated in these models. For multi-field 
DBI inflation, the shape of  non-Gaussianities is found to be the same as in single-field DBI but their amplitude is affected by the transfer between the entropic and adiavatic modes.  The contribution from the scalar field three-point functions  to the coefficient $f_{NL}$ is given by \cite{lrst08a}
\beq
f_{NL}^{(3)}=-\frac{35}{108}\frac{1}{c_s^2}\frac{1}{1+T^2_{{\cal R} {\cal S}} }=-\frac{35}{108}\frac{1}{c_s^2} {\cos^2} \Theta \,.
\label{f_NL3}
\eeq 
 The effect of entropy modes is therefore potentially important in the perspective 
of confronting DBI models to  future CMB observations.

To conclude,  multi-field inflation is a very rich playground, where entropy modes can play a significant role. The most important consequence of entropy modes is the possibility to  modify the curvature perturbation, on large scales, in contrast with single field inflation. This means that the adiabatic fluctuations, which we observe today in the CMB, could come originally from entropy perturbations produced during multi-field inflation.


\vspace{1cm}







\begin{thebibliography}{99}

\bibitem{cargese}
  D.~Langlois,
  ``Inflation, quantum fluctuations and cosmological perturbations,'' in {\it Cargese 2003, Particle physics and cosmology}, p. 235-278 
  [arXiv:hep-th/0405053].
  
\bibitem{wmap5}
  E.~Komatsu {\it et al.}  [WMAP Collaboration],
  arXiv:0803.0547 [astro-ph].
  
\bibitem{Langlois:2005ii}
  D.~Langlois and F.~Vernizzi,
  Phys.\ Rev.\ Lett.\  {\bf 95}, 091303 (2005)
  [arXiv:astro-ph/0503416].
 

\bibitem{Langlois:2005qp}
  D.~Langlois and F.~Vernizzi,
  Phys.\ Rev.\ D {\bf 72}, 103501 (2005)
  [arXiv:astro-ph/0509078].
  
\bibitem{mfb}
V.~F.~Mukhanov, H.~A.~Feldman and R.~H.~Brandenberger,
Phys.\ Rept.\  {\bf 215}, 203 (1992).  
  
\bibitem{Langlois:1999dw}
  D.~Langlois,
  Phys.\ Rev.\ D {\bf 59}, 123512 (1999)
  [arXiv:astro-ph/9906080].

\bibitem{Langlois:2000ar}
  D.~Langlois and A.~Riazuelo,
  Phys.\ Rev.\  D {\bf 62}, 043504 (2000)
  [arXiv:astro-ph/9912497].

\bibitem{Bartolo:2004if}
  N.~Bartolo, E.~Komatsu, S.~Matarrese and A.~Riotto,
  Phys.\ Rept.\  {\bf 402}, 103 (2004)
  [arXiv:astro-ph/0406398].

\bibitem{deltaN}
  M.~Sasaki and E.~D.~Stewart,
  Prog.\ Theor.\ Phys.\  {\bf 95}, 71 (1996)
  [arXiv:astro-ph/9507001];
  A.~A.~Starobinsky,
  JETP Lett.\  {\bf 42}, 152 (1985)
  [Pisma Zh.\ Eksp.\ Teor.\ Fiz.\  {\bf 42}, 124 (1985)].

\bibitem{Lyth:2005fi}
  D.~H.~Lyth and Y.~Rodriguez,
  Phys.\ Rev.\ Lett.\  {\bf 95}, 121302 (2005)
  [arXiv:astro-ph/0504045].


\bibitem{curvaton}
  K.~Enqvist and M.~S.~Sloth,
  Nucl.\ Phys.\  B {\bf 626}, 395 (2002)
  [arXiv:hep-ph/0109214];
  D.~H.~Lyth and D.~Wands,
  Phys.\ Lett.\  B {\bf 524}, 5 (2002)
  [arXiv:hep-ph/0110002];
  T.~Moroi and T.~Takahashi,
  Phys.\ Lett.\  B {\bf 522}, 215 (2001)
  [Erratum-ibid.\  B {\bf 539}, 303 (2002)]
  [arXiv:hep-ph/0110096].
  
\bibitem{Langlois:2004nn}
  D.~Langlois and F.~Vernizzi,
  Phys.\ Rev.\  D {\bf 70}, 063522 (2004)
  [arXiv:astro-ph/0403258].

\bibitem{Lyth:2002my}
  D.~H.~Lyth, C.~Ungarelli and D.~Wands,
  Phys.\ Rev.\  D {\bf 67}, 023503 (2003)
  [arXiv:astro-ph/0208055].
  
\bibitem{lvw08}  
D.~Langlois, F.~Vernizzi and D.~Wands,
  ``Non-linear isocurvature perturbations and non-Gaussianities,''
  arXiv:0809.4646 [astro-ph].
  

\bibitem{ArmendarizPicon:1999rj}
  C.~Armendariz-Picon, T.~Damour and V.~F.~Mukhanov,
  Phys.\ Lett.\  B {\bf 458}, 209 (1999)
  [arXiv:hep-th/9904075].
 
  
  \bibitem{lr08}
  D.~Langlois and S.~Renaux-Petel,
  JCAP {\bf 0804}, 017 (2008)
  [arXiv:0801.1085 [hep-th]].

  \bibitem{lrst08b}
D.~Langlois, S.~Renaux-Petel, D.A.~Steer and T.~Tanaka,
 ``Primordial perturbations and non-Gaussianities in DBI and general multi-field inflation,''
  arXiv:0806.0336 [hep-th].

\bibitem{McAllister:2007bg}
 L.~McAllister and E.~Silverstein,
  Gen.\ Rel.\ Grav.\  {\bf 40} (2008) 565
  [arXiv:0710.2951 [hep-th]].
 
\bibitem{Dvali:1998pa}
  G.~R.~Dvali and S.~H.~H.~Tye,
  Phys.\ Lett.\  B {\bf 450}, 72 (1999)
  [arXiv:hep-ph/9812483].
 

  \bibitem{dbi}
  E.~Silverstein and D.~Tong,
  Phys.\ Rev.\  D {\bf 70}, 103505 (2004)
  [arXiv:hep-th/0310221];
M.~Alishahiha, E.~Silverstein and D.~Tong,
  Phys.\ Rev.\  D {\bf 70}, 123505 (2004)
  [arXiv:hep-th/0404084].
  
\bibitem{lrst08a}
  D.~Langlois, S.~Renaux-Petel, D.~A.~Steer and T.~Tanaka,
  Phys.\ Rev.\ Lett.\  {\bf 101}, 061301 (2008)
  [arXiv:0804.3139 [hep-th]].

\bibitem{Gordon:2000hv}
  C.~Gordon, D.~Wands, B.~A.~Bassett and R.~Maartens,
  Phys.\ Rev.\ D {\bf 63}, 023506 (2001)
  [arXiv:astro-ph/0009131].
 
\bibitem{Garriga:1999vw}
  J.~Garriga and V.~F.~Mukhanov,
  Phys.\ Lett.\  B {\bf 458}, 219 (1999)
  [arXiv:hep-th/9904176].





 \bibitem{sy}
A.~A.~Starobinsky and J.~Yokoyama,
``Density fluctuations in Brans-Dicke inflation'',
gr-qc/9502002


\bibitem{Lalak:2007vi}
  Z.~Lalak, D.~Langlois, S.~Pokorski and K.~Turzynski,
  JCAP {\bf 0707}, 014 (2007)
  [arXiv:0704.0212 [hep-th]].
  
\bibitem{Bartolo:2001rt}
  N.~Bartolo, S.~Matarrese and A.~Riotto,
  Phys.\ Rev.\  D {\bf 64}, 123504 (2001)
  [arXiv:astro-ph/0107502].

\end{thebibliography}
\end{document}